\documentstyle[12pt]{article}
\newtheorem{theorem}{Theorem}[section]

\newcounter{acount}

\newfont{\BB}{msbm10}
\def\R{\mbox{\BB R}}

\def\M{\mbox{\BB M}}
\begin{document}
\title{{\small Revised version 2}\\On the stochastic
mechanics of the free  relativistic particle}
\author{Michele Pavon\\Dipartimento di Elettronica e
Informatica\\Universit\`a di Padova\\and LADSEB-CNR, Italy\\pavon@dei.unipd.it} 
\date{\today}
\maketitle
\newpage
\begin{abstract} Given a positive energy solution of the Klein-Gordon equation, 
the motion of the free, spinless, relativistic particle is described in a fixed Lorentz frame
by a Markov diffusion process with non-constant diffusion coefficient. Proper time is an 
increasing stochastic process and  we derive a probabilistic
generalization of the equation $(d\tau)^2=-\frac{1}{c^2}dX_{\nu}dX_{\nu}$. A
random time-change transformation provides the bridge between the $t$ and the
$\tau$ domain. In the $\tau$ domain, we obtain an $\M^4$-valued
Markov process with singular and constant diffusion coefficient. The square modulus of the 
Klein-Gordon solution is an invariant, non integrable density for this Markov process. It 
satisfies a relativistically covariant continuity equation.
\end{abstract}


{\bf Running title:} Free  relativistic particle
\newpage

\setcounter{equation}{0}
\section {Introduction}
The purpose of this paper is to discuss a possible scenario for the free,
spinless, relativistic particle, extending on Nelson's stochastic mechanics. This
challenging problem has generated considerable interest in the past twenty years, see
\cite{C2,DA,DS,DG,GR,KV,Mo2,MV,N4,S}, and references therein. As is well-known, the main
difficulty in this generalization stems from the non existence of nontrivial
Markov diffusion processes possessing the required relativistic covariance, \cite{DU,HA}.
A number of attempts have been made to circumvent this problem, for instance by considering
different classes of stochastic processes, see e.g. \cite{DA,DS,GR,S}.

In this paper we propose a new approach. Corresponding to a positive energy solution of the Klein-Gordon equation,
motion of the particle is described in a fixed Lorentz frame
by an
$\R^3$-valued Markov diffusion process with {\it non-constant diffusion coefficient} (local
covariance matrix). The
role of {\it proper time} is played here
by an increasing stochastic process, namely a {\em quadratic variation} process. We also derive a natural probabilistic
generalization of the equation $(d\tau)^2=-\frac{1}{c^2}dX_{\nu}dX_{\nu}$. A 
{\it random time-change transformation} provides the bridge between the $t$ and the $\tau$ 
domain. In the $\tau$ domain, we obtain an $\M^4$-valued Markov process with singular and 
constant diffusion coefficient. The fourth component is proportional to a stopping time.
The square modulus of the Klein-Gordon solution is an invariant, non integrable density for this 
Markov process. It satisfies a relativistically covariant continuity equation.

\setcounter{equation}{0}

\section{Background on diffusion processes} In this
section, we review some essential concepts and results of the
kinematics of stochastic mechanics. We refer to
\cite{N1}-
\cite{N3}, \cite {G} for a thorough account. In order to avoid any confusion, stochastic processes
will be denoted by capital letters, as it is customary in probability. Let $(\Omega,{\cal E},{\bf P})$ be a probability
space. A stochastic process
$\{X(t);t_0\le t\le t_1\}$ mapping $[t_0,t_1]$ into
$L^2_n(\Omega,{\cal E},{\bf P})$ is called a {\it finite-energy
diffusion} with constant diffusion coefficient
$I_n\sigma^2$ if the path $X(\omega)$ belongs a.s. to
$C([t_0,t_1];\R^n)$ (n-dimensional continuous functions) and

\begin{equation}\label{K1}
X(t)-X(s)=\int_s^t\beta(\tau)d\tau+\sigma
[W_+(t)-W_+(s)],\quad t_0\le s<t\le t_1, \end{equation} where the
{\it forward drift} $\beta(t)$ is at each time
$t$ a measurable function of the past $\{X(\tau);0\le
\tau\le t\}$, and
$W_+(\cdot)$ is a standard, n-dimensional {\it Wiener process}
with the property that
$W_+(t)-W_+(s)$ is independent of
$\{X(\tau);0\le \tau\le s\}$. Moreover, $\beta$ must satisfy the finite-energy condition
\begin{equation}\label{K2}
E\left\{\int_{t_0}^{t_1}\beta(t)\cdot\beta(t)dt\right\}<\infty.
\end{equation}
In \cite{F}, F\"{o}llmer has shown that a finite-energy diffusion
also admits a reverse-time differential. Namely, there exists a
measurable function $\gamma(t)$ of the future
$\{X(\tau);t\le \tau\le t_1\}$ called {\it backward drift}, and
another Wiener process
$W_-$ such that \begin{equation}\label{K3}
X(t)-X(s)=\int_s^t\gamma(\tau)d\tau+\sigma
[W_-(t)-W_-(s)],\quad t_0\le s<t\le t_1.
\end{equation}  Moreover, $\gamma$ satisfies
\begin{equation}\label{K4}
E\left\{\int_{t_0}^{t_1}\gamma(t)\cdot\gamma(t)dt\right\}<\infty,
\end{equation} and
$W_-(t)-W_-(s)$ is independent of $\{X(\tau);t\le \tau\le t_1\}$. Let us agree that
$dt$ always indicates a strictly positive variable. For any
function $f$ defined on
$[t_0,t_1]$, let
$$d_+f(t):=f(t+dt)-f(t)$$ 
be the {\it forward increment} at time
$t$, and
$$d_-f(t)=f(t)-f(t-dt)$$ be the {\it backward increment} at time
$t$. For a finite-energy diffusion, F\"{o}llmer has also shown in
\cite{F} that the forward and backward drifts may be obtained as
Nelson's conditional derivatives, namely
\begin{equation}\label{D1}\beta(t)=\lim_{dt\searrow
0}E\left\{\frac{d_+X(t)}{dt}|X(\tau), t_0 \le
\tau \le t\right\},\end{equation} and 
\begin{equation}\label{D2}\gamma(t)=\lim_{dt\searrow
0}E\left\{\frac{d_-X(t)}{dt}|X(\tau), t
\le  \tau \le t_1\right\},\end{equation}
the limits being taken in
$L^2_n(\Omega,{\cal B},P)$. It was finally shown in
\cite{F} that the one-time probability density
$\rho(\cdot,t)$ of $X(t)$ (which exists for every $t>t_0$) is absolutely
continuous on
$\R^n$ and the following relation holds $\forall t>0$
\begin{equation}\label{K4'} E\{\beta(t)-\gamma(t)|X(t)\} =
\sigma^2\nabla\log\rho(X(t),t).
\end{equation}
Let us introduce the {\it current drift} $v(t):=(\beta(t)+\gamma(t))/2$ and
the {\it osmotic drift} $u(t):=(\beta(t)-\gamma(t))/2$. Notice that, when $\sigma$
tends to zero, $v$ tends to
$\dot{X},$ and $u$ tends to zero.
The finite-energy diffusion
$X(\cdot)$ is called {\it Markovian} if there exist two
measurable functions
$b_+(\cdot,\cdot)$ and
$b_-(\cdot,\cdot)$ such that
$\beta(t)=b_+(X(t),t)$ a.s. and
$\gamma(t)=b_-(X(t),t)$ a.s., for all $t$ in
$[t_0,t_1]$. The duality relation (\ref{K4'}) now reads
\begin{equation}\label{M1} b_+(X(t),t)-b_-(X(t),t) =
\sigma^2\nabla\log\rho(X(t),t).
\end{equation} This immediately gives the {\it osmotic equation}
\begin{equation}\label{M2}
u(x,t)=\frac{\sigma^2}{2}\nabla\log\rho(x,t),
\end{equation} where $u(x,t):=(b_+(x,t)-b_-(x,t))/2$. The probability density
$\rho(\cdot,\cdot)$ of $X(t)$ satisfies (at least weakly) the
{\it Fokker-Planck equation}
$$
\frac{\partial{\rho}}{\partial{t}} + \nabla
\cdot (b_+\rho) = \frac{\sigma^2}{2}\Delta\rho.
$$  The latter can also be rewritten, in view of (\ref{M1}), as
the {\it equation of continuity} of hydrodynamics
\begin{equation}\label{M3}
\frac{\partial{\rho}}{\partial{t}} + \nabla
\cdot (v\rho) = 0,
\end{equation} where $v(x,t):=(b_+(x,t)+b_-(x,t))/2$.

Nelson's stochastic mechanics \cite{N1,G,N2,BCZ} is a quantization procedure for classical
dynamical systems based on diffusion processes.  Given a quantum
evolution $\{\psi(x,t);t_0\le t\le t_1\}$, namely a solution of the Schr\"{o}dinger equation
\begin{equation}\label{S} \frac{\partial{\psi}}{\partial{t}} =
\frac{i\hbar}{2m}\Delta\psi -
\frac{i}{\hbar}V(x)\psi,
\end{equation}
satisfying Carlen's {\em finite action condition}
\begin{equation}\label{FA}||\nabla\psi||^2_2\in L^1_{{\rm loc}}[[t_0,+\infty)],
\end{equation}
it is possible to construct a measure on path-space under which the canonical 
coordinate process $X$ is a finite-energy Markov diffusion process.  Writing $\psi(x,t) = \exp
{R(x,t) + \frac{i}{\hbar} S(x,t)}$, we have that
the process $X$ has current and 
osmotic drift fields given, respectively, by $v(x,t) =\frac{1}{m}\nabla{S(x,t)}$, and 
$u(x,t) =\frac{\hbar}{m}\nabla{R(x,t)}$. In particular, the (forward) Ito differential
of $X$ is given by
\begin{eqnarray}\nonumber dX(t)&=&\left[\frac{\hbar}{m}\nabla \left(\Re
\log\psi(X(t),t)
+\Im \log\psi(X(t),t)\right)\right]dt\\
\label{N}
&+&\sqrt{\frac{\hbar}{m}}dW_+(t),
\end{eqnarray}
see \cite{C}, \cite [Chapter IV]{BCZ}, and references therein.
Moreover, the
probability density $\rho(\cdot,t)$ of
$x(t)$ satisfies
\begin{equation}\label{D}\rho(x,t)=|\psi(x,t)|^2,\quad \forall t \in [t_0,t_1].
\end{equation}
We now need to recall the generalization to Markov processes with diffusion coefficient of the form
$I_n\sigma(x,t)^2$, with $\sigma(x,t)\in\R$, the above kinematics, see
e.g.
\cite{N0}, \cite{Na}, \cite{Mo} and \cite{HP}. Let
$\{X(t);t_0\le t\le t_1\}$ be an $n$-dimensional Markov
diffusion process whose increments admit the two representations
\begin{eqnarray}\label{QR14}d_+X(t)=b_+(X(t),t)dt+\sigma_+(X(t),t)d_+W_+(t),\\
d_-X(t)=b_-(X(t),t)dt+\sigma_-(X(t),t)d_-W_-(t),\label{QR15}
\end{eqnarray} where the drifts and diffusion coefficients are
sufficiently regular. Here, $W_+$ and
$W_-$ are standard, n-dimensional Wiener processes adapted to the
past and the future, respectively, of the process $X$. Then,
the following relations hold
\begin{eqnarray}\sigma_+(x,t)=\sigma_-(x,t)=\sigma(x,t)\\
 b_+(x,t)-b_-(x,t) =\frac{1}{\rho}
\nabla\left(\sigma(x,t)^2\rho(x,t)\right).\label{QQ}
\end{eqnarray} Letting, as usual, $v:=(b_++b_-)/2$ denote
the current drift, we have, as before, that the Fokker-Planck
equation can be rewritten as the continuity equation (\ref{M3}). Let
$R(x,t):=\frac{1}{2}\log\left(\sigma(x,t)^2\rho(x,t)\right)$.
Then, observing that (\ref{QQ}) can be rewritten as
$$b_+(x,t)-b_-(x,t) =\sigma(x,t)^2
\nabla\log\left(\sigma(x,t)^2\rho(x,t)\right),$$ we get the
following expression for the osmotic drift $u$
\begin{equation}\label{QR17} u(x,t)=\sigma(x,t)^2\nabla R(x,t).
\end{equation}

\setcounter{equation}{0}
\section{Description in a fixed inertial frame}

For the definitions
and  results about continuous martingales that occur in this section, we refer the reader
to the Appendix and, for a thorough treatment, to
\cite{KS} and
\cite{LS}. Let
$\varphi:=\exp[R+\frac{i}{\hbar}S]$ be a solution of the Klein-Gordon equation
\begin{equation}\label{KG}\Delta\varphi-\frac{1}{c^2}\frac{\partial^2}{\partial
t^2}\varphi=\frac{m^2c^2}{\hbar^2}\varphi
\end{equation}
on $[0,\infty)$, such that $S$ satisfies on the same time interval
\begin{equation}\label{Z0}			
\frac{\partial{S}}{\partial{t}} + c\sqrt{m^2c^2+\nabla
S\cdot\nabla S} = 0.
\end{equation}
This is precisely the family of solutions that was considered in \cite{MV}. Indeed, let
\begin{eqnarray}\label{rho}\rho(x,t)=\frac{i\hbar}{2mc^2}\left(\bar{\varphi}\frac{\partial \varphi}{\partial t}-
\varphi\frac{\partial \bar{\varphi}}{\partial t}\right)(x,t)=
|\varphi(x,t)|^2\times \left(-\frac{1}{mc^2}\frac{\partial
S}{\partial t}(x,t)\right),\\{\bf j}(x,t)=\frac{\hbar}{2mi}\left(\bar{\varphi}
\nabla\varphi-
\varphi\nabla\bar{\varphi}\right)(x,t)=|\varphi(x,t)|^2\times \left(\frac{1}{m}\nabla
S(x,t)\right).
\end{eqnarray}
First of all, observe that because of (\ref{KG}), the pair $(\rho,{\bf j})$ satisfies a continuity type equation
$$\frac{\partial{\rho}}{\partial{t}}+\nabla\cdot{\bf j}=0.
$$
Moreover, from (\ref{Z0}) it also follows that
\begin{eqnarray}
\rho\ge 0,\\{\bf j}\cdot{\bf j}-c^2\rho^2=-c^2|\varphi|^4.
\end{eqnarray}
The latter two conditions are equivalent to $(5.10)$ in \cite[p.4705]{MV}. 
This class of solutions is non empty. It contains at least all 
positive-frequency plane-wave solutions, namely solutions of the form
$$\varphi(x,t)=\exp [(i/\hbar)({\bf p}\cdot x-p^0t)],
$$
where
$$p^0=+c\sqrt{{\bf p}\cdot{\bf p}+m^2c^2}.
$$   
Let
$$\rho_0(x):=\frac{i\hbar}{2mc^2}\left(\bar{\varphi}\frac{\partial \varphi}{\partial t}-
\varphi\frac{\partial \bar{\varphi}}{\partial t}\right)(x,0)=
|\varphi(x,0)|^2\times \left(-\frac{1}{mc^2}\frac{\partial
S}{\partial t}(x,0)\right).$$
It follows from (\ref{Z0}) that $\rho_0(x)\ge 0$.
Let us suppose that $\rho_0$ integrates to one. Suppose that $\{X(t);
t\ge 0\}$ is a Markov diffusion process such that $X(0)$ is
distributed according to $\rho_0$, and having forward differential
\begin{equation}\label{Z1} d_+X(t)=\left(\frac{\frac{1}{m}\nabla
S+\frac{\hbar}{m}\nabla R}{-\frac{1}{mc^2}\frac{\partial
S}{\partial
t}}\right)(X(t),t)dt+
\sqrt{\frac{\hbar}{m}}\frac{1}{\sqrt{-\frac{1}{mc^2}\frac{\partial
S}{\partial t}(X(t),t)}}d_+W_+(t),
\end{equation}
where $W_+(\cdot)$ is a standard, 3-dimensional Wiener process. In the previous notation,
we have
$$\sigma^2(x,t):=\frac{\hbar}{-\frac{1}{c^2}\frac{\partial
S}{\partial t}(x,t)}.$$
A standard calculation \cite{Sch} shows that
$\rho(x,t)$ in (\ref{rho}) satisfies the Fokker-Planck equation corresponding to (\ref{Z1}). Hence, $\rho(x,t)$ is the
probability density of $X(t)$. Moreover, the current drift of $X$ is 
\begin{equation}\label{Z1'}v(X(t),t):=\left(\frac{\frac{1}{m}\nabla
S}{-\frac{1}{mc^2}\frac{\partial S}{\partial
t}}\right)(X(t),t),
\end{equation}
and the osmotic drift is given by
\begin{equation}\label{Z1''}
u(X(t),t)=\left(\frac{\frac{\hbar}{m}\nabla R}{-\frac{1}{mc^2}\frac{\partial
S}{\partial
t}}\right)(X(t),t).
\end{equation}

Let us
define
\begin{equation}\label{Z2}
M^i(t):=\int_{0}^{t}\frac{1}{\sqrt{-\frac{1}{mc^2}\frac{\partial
S}{\partial s}(X(s),s)}}d_+W_+^i(s),\quad t\ge 0\quad i=1,2,3.
\end{equation}
The stochastic processes $\{M^i(t);t\ge 0\}$ are continuous {\it
local martingales} with the same {\it quadratic variation}
\begin{equation}\label{Z3}
<M>(t)=<M^i>(t):=\int_{0}^{t}\frac{1}{-\frac{1}{mc^2}\frac{\partial
S}{\partial s}(X(s),s)}ds.
\end{equation}
Moreover, their {\it cross variations} satisfy
\begin{equation}\label{Z3'}
<M^i,M^j>(t)=0,\quad 1\le i\neq j\le 3,\quad \forall t\ge 0.
\end{equation}
Observe that $<M>(t)$ is a strictly increasing process with
differentiable sample paths. It satisfies, in view of (\ref{Z0}) and (\ref{Z1'}),
\begin{eqnarray}\nonumber
d<M>(t)&=&\frac{1}{-\frac{1}{mc^2}\frac{\partial
S}{\partial t}(X(t),t)}dt\\\nonumber&=&\sqrt{\frac{m^2c^2}{m^2c^2+\nabla S(X(t),t)\cdot\nabla
S(X(t),t)}}dt\\&=&\sqrt{1-\frac{\nabla S(X(t),t)\cdot\nabla
S(X(t),t)}{m^2c^2+\nabla S(X(t),t)\cdot\nabla
S(X(t),t)}}dt\nonumber\\&=&\sqrt{1-\frac{v(X(t),t)\cdot v(X(t),t)}{c^2}}dt.\label{PT}
\end{eqnarray}
If for some $t$ and $\omega$ we have
$v(X(t,\omega),t)=0$, then $d<M>(t)=dt$. We can therefore think of
$d<M>(t)$ as of an increment of the particle {\it random proper time}. Notice that (\ref{PT})
implies
$$\frac{d<M>(t)}{dt}\le 1,\quad a.s.
$$
Hence,
\begin{equation}\label{MO}
<M>(t)\le t,\quad a.s.,\quad \forall t\ge 0.
\end{equation}
Furthermore, we have the following property of
$d<M>(t)$. Let us first recall that, in view of (\ref{D1}) and (\ref{D2}), we have
\begin{equation}\label{D3}v(X(t),t)=\lim_{dt\searrow
0}E\left\{\frac{d_sX(t)}{dt}|X(t)\right\},
\end{equation}
where the {\it symmetric increment} of $X$ at time $t$ is defined by
$$d_sX(t):=\frac{d_+X(t)+d_-X(t)}{2}=\frac{X(t+dt)-X(t-dt)}{2}.$$ We can therefore
rewrite (\ref{PT}) as follows.
\begin{eqnarray}\nonumber c^2\left[d<M>(t)\right]^2=-\left[(ic)^2+v(X(t),t)\cdot
v(X(t),t)\right](dt)^2\\\nonumber=-\left[(\frac{d(ict)}{dt})^2+v(X(t),t)\cdot
v(X(t),t)\right](dt)^2\\=-\left[(d(ict))^2+E\{d_sX(t)|X(t)\}\cdot
E\{d_sX(t)|X(t)\}+o((dt)^2)\right].\label{PT2}
\end{eqnarray}
Let us introduce
$$X_{\nu}(t)=\left(\begin{array}{c}X(t)\\ict\end{array}\right).$$
Then (\ref{PT2}) can be written as follows
\begin{equation}\label{PT3}\left[d<M>(t)\right]^2=-\frac{1}{c^2}E\{d_sX_{\nu}(t)|X_{\nu}(t)\}\cdot
E\{d_sX_{\nu}(t)|X_{\nu}(t)\} +o((dt)^2),
\end{equation}
generalizing the relation $(d\tau)^2=-\frac{1}{c^2}dX_{\nu}dX_{\nu}$ of 
classical relativistic mechanics. 

\setcounter{equation}{0}
\section{From the $t$ to the $\tau$ formulation: A random time-change}

An important consequence of Levy's characterization of
the Wiener process, see e. g. \cite[p. 82]{LS}, is that any continuous local
martingale may be viewed as a time-changed Wiener process, cf. Theorem A.2 in the Appendix. For
$\tau\ge 0$, let us introduce the {\it stopping time}
$$T(\tau):=\inf\{\sigma\ge 0:<M>(\sigma)\ge \tau\}.
$$
(In the case when the probability of the event
$\{\omega:\lim_{t\rightarrow\infty}<M>(t,\omega)=\infty\}$ is strictly less
than one, the stopping time $T(\tau)$ has to be suitably modified, see
\cite[pp.174-175]{KS}). Notice that  \begin{equation}\label{Z4}<M>(T(\tau))=\tau,
\quad T(<M>(t))=t. \end{equation} Then, the processes
$$\tilde{W}_+^i(\tau):=M^i_{T(\tau)}, \quad i=1,2,3,$$
are, standard, one-dimensional Wiener processes. Moreover, in view of
property (\ref{Z3'}), we can apply a theorem of F. B. Knight (see Theorem A.3 in the Appendix),
and conclude that the processes
$\tilde{W}_+^i(\tau)$
are pairwise independent, and can, consequently, be viewed as the
components of a standard, three-dimensional Wiener
process $\tilde{W}_+(\tau)$. Let us introduce the stochastic process 
$$\tilde{X}(\tau):=X(T(\tau)), \quad \tau\ge 0.$$ In view of (\ref{Z1}),
(\ref{Z3}), and (\ref{Z4}), the forward differential of $\tilde{X}$ is given by
\begin{equation}\label{Z5}
d_+\tilde{X}(\tau)=\left(\frac{1}{m}\nabla S+\frac{\hbar}{m}\nabla
R\right)(\tilde{X}(\tau),T(\tau))d\tau+\sqrt{\frac{\hbar}{m}}d_+\tilde{W}_+(\tau).
\end{equation}
Notice that $\tilde{X}(\tau)$ is a non-Markovian $\R^3$-valued diffusion process with
constant diffusion coefficient $\frac{\hbar}{m}I_3$. The Markov property has
therefore been destroyed by the random time change. From the first relation in (\ref{Z4}),
we get
$$\frac{dT(\tau)}{d\tau}=\left[\frac{d<M>}{dt}(T(\tau))\right]^{-1}.$$
In view of (\ref{Z3}), we then get
\begin{equation}\label{Z6}
\frac{dT(\tau)}{d\tau}=-\frac{1}{mc^2}\frac{\partial
S}{\partial t}(\tilde{X}(\tau),T(\tau)).
\end{equation}
Let us now define $\tilde{X}^4(\tau):=icT(\tau)$, $\tilde{S}(x_\nu)=\tilde{S}(x,x^4):=S(x,t)$,
and 
$\tilde{R}(x_\nu)=\tilde{R}(x,x^4):=R(x,t)$. Equations (\ref{Z5}) and (\ref{Z6}) can now be
rewritten as
\begin{eqnarray}\label{Z7}
d_+\tilde{X}(\tau)&=&\left(\frac{1}{m}\nabla \tilde{S}+\frac{\hbar}{m}\nabla
\tilde{R}\right)(\tilde{X}(\tau),\tilde{X}^4(\tau))d\tau+\sqrt{\frac{\hbar}{m}}d_
+\tilde{W}_+(\tau)
\\\label{Z8}
d_+\tilde{X}^4(\tau)&=&\frac{1}{m}\frac{\partial \tilde{S}}{\partial
x^4}(\tilde{X}(\tau),\tilde{X}^4(\tau))d\tau.
\end{eqnarray}
Notice that $\tilde{X}_{\nu}(\tau)$ defined by
$$ 
\tilde{X}_{\nu}(\tau)=\left(\begin{array}{c}\tilde{X}(\tau)\\\tilde{X}_4(\tau)\end{array}\right).$$
is an $\M^4$-valued {\it Markovian} stochastic process with 
marginal density at $\tau=0$ given by $\rho_0(x)\cdot\delta(x^4)$ since $T(0)=0$ a.s. 
Let us introduce the diffusion matrix
$$\Sigma^2:=\left(\begin{array}{cc}\frac{\hbar}{m}I_3&0\\0&0\end{array}\right)$$. Then 
(\ref{Z7})-(\ref{Z8})
can be rewritten in the form
\begin{equation}\label{INV}d_+\tilde{X}_\nu(\tau)=\left(\frac{1}{m}\nabla_\nu \tilde{S}+\Sigma^2\nabla_\nu
\tilde{R}\right)(\tilde{X}_\nu(\tau))d\tau+\Sigma d_
+\tilde{W}_\nu(\tau),
\end{equation}
where $\tilde{W}_\nu$ is any standard, four-dimensional Wiener process whose first 
three components form $\tilde{W}_+$. The Fokker-Plank equation is then
\begin{equation}\label{FPM}
\frac{\partial{\rho_\nu}}{\partial{\tau}} + \nabla_\nu
\cdot \left[\left(\frac{1}{m}\nabla_\nu \tilde{S}+\Sigma^2\nabla_\nu
\tilde{R}\right)\rho_\nu\right] = \frac{1}{2}\Sigma^2\Delta_\nu\rho_\nu.
\end{equation}
Because of (\ref{KG}),
$\varphi(x_{\nu}):=\exp[\tilde{R}(x_{\nu})+\frac{i}{\hbar}\tilde{S}(x_\nu)]$ satisfies 
$$
\nabla_{\nu}\cdot\nabla_{\nu}\varphi=\frac{m^2c^2}{\hbar^2}\varphi.
$$
The latter is equivalent to the system of p.d.e.'s
\begin{eqnarray}\label{A11}
-\frac{1}{2m}\nabla_{\nu}\tilde{S}\cdot\nabla_{\nu}\tilde{S}+
\frac{\hbar^2}{2m}\left[\nabla_{\nu}\tilde{R}\cdot\nabla_{\nu}\tilde{R}+
\nabla_{\nu}\cdot\nabla_{\nu}\tilde{R}\right]&=&\frac{mc^2}{2},\\\label{A12}
\frac{1}{m}\nabla_{\nu}\tilde{S}\cdot\nabla_{\nu}\tilde{R}+
\frac{1}{2m}\nabla_{\nu}\cdot\nabla_{\nu}\tilde{S}&=&0.
\end{eqnarray} 
Let $$\tilde{\rho}(x_{\nu})=|\varphi(x_\nu)|^2.$$
From (\ref{A12}), we get that $\tilde{\rho}(x_{\nu})$ satisfies (\ref{FPM}).
Thus, $\mu(dx_\nu):=\tilde{\rho}(x_{\nu})dx_\nu$ is an {\em invariant  ($\sigma$-finite) measure} 
for (\ref{Z7})-(\ref{Z8}). Notice that $\tilde{\rho}(x_{\nu})$ also satisfies the 
manifestly covariant equation
$$\nabla_{\nu}\cdot(\tilde{\rho}\frac{1}{m}\nabla_{\nu}\tilde{S})=0.
$$
If we give  
$\mu(dx_\nu):=\tilde{\rho}(x_{\nu})dx_\nu$ as initial measure to (\ref{INV}), the 
probabilistic picture is lost. Nevertheless, it is possible to make sense of the 
time-reversed diffusion along the lines of \cite[pp.44-45]{BCZ}. Moreover, consider the 
{\em generator} of $\tilde{X}_{\nu}$ acting on smooth functions with compact support
$$\left(\frac{1}{m}\nabla_\nu \tilde{S}+\Sigma^2\nabla_\nu
\tilde{R}\right)\cdot\nabla_\nu+\frac{1}{2}\Sigma^2 \Delta_\nu.$$
The adjoint of this operator with respect to the measure $\mu(dx_\nu)$ \cite[p.104]{N1} is given by
$$\left(\frac{1}{m}\nabla_\nu \tilde{S}-\Sigma^2\nabla_\nu
\tilde{R}\right)\cdot\nabla_\nu-\frac{1}{2}\Sigma^2 \Delta_\nu.
$$
Hence, in spite of the lack of a probabilistic picture we can still think of 
$\frac{1}{m}\nabla_\nu \tilde{S}$ as of a bilateral velocity field associated to 
the equilibrium measure $\mu(x_\nu)$. Then, recalling that (\ref{Z0}) implies
$$\nabla_{\nu}\tilde{S}(x_{\nu})\cdot\nabla_{\nu}\tilde{S}(x_{\nu})=-m^2c^2,
$$
we get a $\tau$-domain counterpart of relation (\ref{PT3}).

\section{Conclusion and outlook}
Although the process (\ref{Z7})-(\ref{Z8}) is not relativistically covariant, our description 
does not appear to be in conflict with classical relativistic mechanics nor with Nelson's 
nonrelativistic theory. Indeed, in the case when we let 
the noise intensity in (\ref{Z1}) tend to zero, we recover the equations of 
classical special relativity. In particular, the quadratic variation $<M>$ 
defined by (\ref{Z3}) tends to proper time $\tau$ and $T(\tau)$ tends to $t(\tau)$. 
On the other hand, when $v\cdot v<< c^2$, 
motion of the particle in a fixed inertial frame is described by a Markov diffusion 
process with diffusion coefficient nearly equal to $\frac{\hbar}{m}$.  The Schr\"odinger 
equation is then recovered through some appropriate procedure neglecting terms multiplied by 
$\hbar^2/c$, cf. \cite[p. 4708]{MV}.

\vspace{0.25cm}

{\bf Acknowledgments:} The author wishes to thank Paolo Dai Pra
for some helpful comments on the results of this paper, and
Lorenza Viola for several useful conversations on relativistic
stochastic mechanics.

\appendix
\setcounter{equation}{0}
\section{Continuous Martingales}
We collect in this appendix a few basic
facts about continuous martingales. We refer the reader to \cite{KS} and \cite{LS} 
for the proofs and more information.
Let $(\Omega,{\cal A},{\bf P})$ be a probability space, and let ${\cal
F}:=({\cal F}_t)$, $t\in [t_0,t_1]$, be a non-decreasing family of sub $\sigma$
-algebras of ${\cal A}$. It will be always assumed that ${\cal F}_{t_0}$ contains all the
zero-probability sets in ${\cal A}$, and that the filtration is right-continuous, namely
$\{\cap{\cal F}_t, t>s\}={\cal F}_s$, for every $s$. A stochastic process
$X:=\{X(t);t\in [t_0,t_1]\}$ is said to be {\it continuous} if it has continuous trajectories
$X(t,\omega),t\in [t_0,t_1]$, with probability one. It is called ${\cal F}$-{\it adapted} if, for
every
$t$ in
$[t_0,t_1]$, $X(t)$ is
${\cal F}_t$-measurable. In that case, it is customary to write $X=(X(t),{\cal F}_t)_{t\in
[t_0,t_1]}$. The stochastic process
$M=(M(t),{\cal F}_t)_{t\in [t_0,t_1]}$ is called a {\it martingale} (with respect to the {\it
filtration} ${\cal F}_t,t\in [t_0,t_1]$) if it satisfies the two following conditions:
\begin{enumerate}
\item $E\left\{|M(t)|\right\}<\infty,\quad \forall t$,
\item $E\left\{M(t)|{\cal F}_s\right\}=M(s),\quad t\ge s$.
\end{enumerate}
The continuous martingale $X=(X(t),{\cal F}_t)_{t\in
[t_0,t_1]}$ is called {\it square-integrable} if $E\left\{X(t)^2\right\}<\infty,\quad \forall t$ 
in $[t_0,t_1]$. Let $X=(X(t),{\cal F}_t)_{t\in
[t_0,t_1]}$ be a continuous, square-integrable martingale with $X(t_0)=0$ with probability
one. Then, the celebrated Doob-Meyer decomposition theorem implies that there exists a unique
representation for $X(t)^2$
$$X(t)^2=<X>(t)+M(t),\quad t\in [t_0,t_1],
$$
where $<X>$ is an adapted, continuous, increasing process with $<X>(t_0)=0$, and $M$ is a continuous
martingale with $M(t_0)=0$. The process $<X>(\cdot)$ is called the {\it quadratic variation} of
$X(\cdot)$. Consider now two continuous, square-integrable martingales $X=(X(t),{\cal F}_t)_{t\in
[t_0,t_1]}$ and $Y=(Y(t),{\cal F}_t)_{t\in
[t_0,t_1]}$. Their {\it cross-variation process} $<X,Y>$ is defined for $t\in [t_0,t_1]$ by
$$<X,Y>(t):=\frac{1}{4}\left[<X+Y>(t)-<X-Y>(t)\right].
$$
The cross variation of $X$ and $Y$ is characterized by the fact that $XY-<X,Y>$ is a continuous
martingale.

A nonnegative random variable $T(\omega)$ is called a {\it Markov time} or a {\it stopping
time} (relative to the {\it
filtration} ${\cal F}_t,t\in [t_0,t_1]$) if, for all $t$ in $[t_0,t_1]$,
$$\{\omega:T(\omega)\le t\}\in {\cal F}_t. 
$$
Let $T$ be a stopping time relative to ${\cal F}_t,t\in [t_0,t_1]$. Then, the $\sigma$-field 
${\cal F}_T$ {\it of events determined prior to the stopping time} $T$ consists of those events
$A\in {\cal A}$ for which $\left[A\cap \{\omega: T(\omega)\le t\}\right]\in {\cal F}_t$, for every
$t$ in $[t_0,t_1]$.

The stochastic process
$M=(M(t),{\cal F}_t)_{t\ge 0}$, is called a {\it local martingale} if there exists an increasing
sequence of stopping times (with respect to
${\cal F}$)
$\{T_n\},n=1,2,\ldots$ such that
\begin{enumerate}
\item ${\bf P}(\lim T_n=\infty)=1$;
\item for any $n$, the ``stopped process" $(M(t\wedge T_n), t\ge 0$, is a martingale.	
\end{enumerate}
The decomposition result for the product
of two square-integrable martingale extends to local martigales as follows. Let $X=(X(t),{\cal
F}_t)_{t\in [t_0,t_1]}$ and $Y=(Y(t),{\cal F}_t)_{t\in [t_0,t_1]}$ be two continuous, local
martingales. Then there is a unique adapted process $<X,Y>$ such that $XY-<X,Y>$ is a
continuous local martingale. We write $<X>$ instead of $<X,X>$. We can now state P. Levy's
martingale characterization of the Wiener process.
\begin{theorem}
 Let $M=(M(t),{\cal F}_t)_{t\ge 0}$ be a continuous local martingale with $M(0)=0$ a.s.
and quadratic variation $<M>(t)=t$. Then $M$ is a standard Wiener process. 
\end{theorem}
As a corollary to this theorem, we have the following result showing that continuous local
martingales can be viewed as time-changed Wiener processes.
\begin{theorem} Let  $M=(M(t),{\cal F}_t)_{t\ge 0}$ be a continuous local martingale with $M(0)=0$
a.s. Suppose that $\lim_{t\rightarrow\infty}<M>(t)=\infty$ with probability one. Define, for
$\tau\ge 0$, the stopping time $T(\tau):=\inf\{t\ge 0:<M>(t)>\tau\}$. Then the time-changed
process $W=\{W(\tau):=M(T(\tau)),{\cal G}_{\tau}:={\cal F}_{T(\tau)})_{\tau\ge 0}$ is a standard
Wiener process and we have, with probability one,
$$M(t)=W(<M>(t)),\quad t\ge 0.
$$
\end{theorem}
The multivariate extension of this theorem is due to F. B. Knight \cite[p.179]{KS}.
\begin{theorem}Let  $M=\{M(t)=(M^1(t),\ldots,M^n(t)),{\cal F}_t)_{t\ge 0}\}$ be a continuous
process with $M(0)=0$ a.s. Suppose that the components $M^i$ are local martingales satisfying the
two following conditions:
\begin{enumerate}
\item $\lim_{t\rightarrow \infty}<M^i>(t)=\infty,\quad i=1,\ldots,n$ a.s.;
\item $<M^i,M^j>(t)=0,\quad i\neq j, t\ge 0$.
\end{enumerate}
Let $T_i(\tau):=\inf\{t\ge 0:<M^i>(t)>\tau\}, t\ge 0, i=1,\ldots,n$. Then the $W^i(\tau):=M^i 
(T_i(\tau)), \tau\ge 0, i=1,\ldots,n$ are {\rm independent}, standard Wiener processes.
\end{theorem}
The nontrivial content of this theorem is that, although the $M^i$ are not independent, applying
first the appropriate time-changes, and then forgetting the time changes, we get independent Wiener
processes. Forgetting the time-changes consists in replacing the filtrations $\{{\cal
G}^i_{\tau}\}$ with the poorer filtrations $\{{\cal F}_{\tau}^{W^i}\}$.

\end{document}